# High-Performance Computing for Scheduling Decision Support: A Parallel Depth-First Search Heuristic

## Gerhard Rauchecker
Department of Management Information Systems
University of Regensburg
Germany
Email: gerhard.rauchecker@wiwi.uni-regensburg.de

## Guido Schryen
Department of Management Information Systems
University of Regensburg
Germany
Email: guido.schryen@wiwi.uni-regensburg.de

## Abstract

Many academic disciplines - including information systems, computer science, and operations management - face scheduling problems as important decision making tasks. Since many scheduling problems are NP-hard in the strong sense, there is a need for developing solution heuristics. For scheduling problems with setup times on unrelated parallel machines, there is limited research on solution methods and to the best of our knowledge, parallel computer architectures have not yet been taken advantage of. We address this gap by proposing and implementing a new solution heuristic and by testing different parallelization strategies. In our computational experiments, we show that our heuristic calculates near-optimal solutions even for large instances and that computing time can be reduced substantially by our parallelization approach.

## 1    Introduction

Scheduling problems can be found in several academic disciplines. For example, in cloud computing, applications are scheduled (Beloglazov et al. 2012; van der Meer et al. 2012; Yang et al. 2014), in energy management, enterprise resources have to be scheduled (Bodenstein et al. 2011; Brandt and Bodenstein 2012; Zhao et al. 2013), and in sports management, tournaments and leagues are scheduled (Duran et al. 2014; Nurmi et al. 2014; Su et al. 2013). In healthcare management, appointments and visits are scheduled (Mak et al. 2015; Meyer et al. 2014; Paulussen et al. 2013), in disaster management, rescue units are scheduled (Rolland et al. 2010; Schryen et al. 2015; Wex et al. 2014; Wex et al. 2011), and in information systems, scheduling systems are built (Chandra et al. 2012; Faghihi et al. 2014; Miranda et al. 2012). In computer science, software application jobs are assigned to computer processors (Li and Peng 2011; Silberschatz et al. 2013; Terekhov et al. 2014), in operations management, production jobs are scheduled on machines (Joo and Kim 2015; Mensendiek et al. 2015; Wang and Liu 2013) and workers are assigned to shifts or jobs (Cote et al. 2011; Elahipanah et al. 2013; Rauchecker et al. 2014), and in logistics, transportation scheduling problems occur (Emde and Boysen 2014; Sterzik and Kopfer 2013; Ullrich 2013).

As numerous scheduling problems are NP-hard (Pinedo 2012), which means that there is no algorithm that can solve the problem to optimality in polynomial time, many large real-world instances are computationally intractable due to time limitations. Thus, there is a need for heuristics which are computationally efficient. However, heuristics lead to suboptimal decisions which result in a waste of time, increased costs, and even fatalities. Therefore, it is important for solution heuristics not only to run computationally efficient but also to be effective, i.e., to find near-optimal solutions. While the effectiveness of (deterministic) heuristics is fixed by their algorithms, the efficiency can often be increased by using modern parallel hardware architectures.

Recent developments in high performance computing (HPC) have led to a substantial increase in computing power. For example, modern PCs and even smartphones have multiple cores, which allow for parallel code execution. At the extreme, computer clusters and supercomputers, which have up to several hundreds of thousands or even several millions of cores (TOP500 2014), are pushing the boundaries of HPC. Supercomputers have shown an exponential growth in peak performance, with the Tianhe-2 (MilkyWay-2), currently the fastest supercomputer, having more than 54 PFlop/s (5.4·10$^{16}$ floating point operations per second) (TOP500 2014). However, HPC does not require access to a supercomputer; it can also be done on computing clusters, which have become commodity IT resources. For example, they are available at many universities and are provided by some cloud providers, especially by Amazon Web Services (Mauch et al. 2013). To sum up, HPC has not only become technologically feasible, but also economically affordable.

In this paper, we focus on a specific problem which occurs in many application domains: the parallel machine scheduling problem on m unrelated machines, sequence- and machine-dependent setup times, machine eligibility restrictions, and a total weighted completion time objective function, classified by Pinedo (2012) as $Rm/s_{ijk}, M_j/\sum w_j C_j$ in the established α/β/γ-notation (Graham et al. 1979) and proven to be NP-hard by Wex et al. (2014).

For this scheduling problem, we address two research questions:

- How can near-optimal solutions for large instances of the scheduling decision problem be found in reasonable time?
- How efficient is the application of HPC to the scheduling decision problem?

To answer these questions, we propose a heuristic based on an exact branch-and-price (b&p) algorithm, which was originally formulated by Lopes and de Carvalho (2007), and evaluate a parallel implementation of the heuristic on a high performance cluster. We demonstrate the efficiency of our heuristic in terms of runtime and the performance of its parallelization based on an established scalability metric. Further, we show the effectiveness of our heuristic by (1) reporting upper bounds of the gap between the (unknown) optimal solution and the heuristic solution and (2) comparing it with an established heuristic.

The remainder of this paper is structured as follows: We present the literature related to our research from both the scheduling and the HPC perspective in the following section. The third section outlines the mathematical model of our scheduling problem, a sophisticated branch-and-price algorithm based heuristic, and our techniques to develop a parallel implementation of the heuristic. In the fourth section, we present our computational experiments before presenting and discussing our results in section five and closing the article with a conclusion.

## 2   Literature Review

In this section, we present the relevant literature for our approach from different perspectives. We outline achievements and limitations of existing works in each subsection, which leads to formulating the research questions proposed in the introduction.

### 2.1   Scheduling Decision Support

Scheduling problems appear in many forms and have attracted thousands of research papers which deal with different solution methods to support decision making in real-world settings. In order to structure this large body of research, several comprehensive literature reviews have been conducted. In their well-established surveys, Allahverdi et al. (1999; 2008) classify scheduling problems into those which account for setup times (costs) and those which do not. Problems of the former type are further classified along the dimensions single machine/parallel machines, batch/non-batch and sequence-dependent/sequence-independent setup times. Using this classification, the scheduling problem $Rm/s_{ijk}, M_j/\sum w_j C_j$, being considered in this paper, is a generalization of the problem class parallel-machine scheduling on unrelated machines, non-batch sequence-dependent setup times, and a total weighted completion time objective function. To be more precise, the problem class formulated in the literature is more restrictive than our problem ($Rm/s_{ijk}, M_j/\sum w_j C_j$) in the sense that, in the former problem class, setup times are machine-independent and each machine is capable of processing each job.

### 2.2   Exact Algorithms

Regarding the NP-hard scheduling problem $Rm/s_{ijk}, M_j/\sum w_j C_j$ and related types of problems, there are only a few research papers that present exact solution algorithms. The survey of Li and Yang (2009) lists two articles with exact solution algorithms for the problem $Rm/\sum w_j C_j$ (which does not account for setup times). The algorithms are capable of solving small instances with 25 jobs and 2 machines in less than 15 minutes (Azizoglu and Kirca 1999) and medium-sized instances with 100 jobs and 20 machines in less than one hour (Chen and Powell 1999). Lopes and de Carvalho (2007) presented a b&p strategy for the parallel machine scheduling problem $Rm/s_{ij}, r_j, d_j/\sum w_j T_j$ on unrelated machines with sequence- (but not machine-) dependent setup times, release dates, due dates, and a total weighted tardiness objective function. Their algorithm is capable of solving instances with 150 jobs and 50 machines in less than one hour.

Another approach is to model our scheduling problem as a quadratic binary program and to have it solved using off-the-shelf optimization software, such as GUROBI or CPLEX. However, computational studies indicate that this strategy is inefficient as it fails to compute optimal solutions for small-sized instances consisting of 40 jobs and 10 machines within several hours (Schryen et al. 2015).

In summary, exact algorithms for problems similar to $Rm/s_{ijk}, M_j/\sum w_j C_j$, such as $Rm/\sum w_j C_j$ and $Rm/s_{ij}, r_j, d_j/\sum w_j T_j$, have been scarcely addressed in the literature and are not capable of solving large instance sizes in reasonable time.

## 2.3 Heuristics

Regarding parallel machine scheduling on unrelated machines with non-batch sequence-dependent setup times, there are only a few publications that develop solution heuristics (Lin and Ying 2014). Kim et al. (2002) and Low (2005) use Simulated Annealing to minimize the total tardiness and the total flow time, while Vallada and Ruiz (2011) minimize the makespan using a genetic algorithm. A Tabu Search to minimize total tardiness, total weighted tardiness, maximum tardiness, and maximum lateness has been investigated by Chen (2006), Chen and Wu (2006), Kim and Shin (2003), and Logendran et al. (2007). Rabadi et al. (2006) introduced a Randomized Priority Search metaheuristic to minimize the makespan while de Paula et al. (2007) presented a Variable Neighborhood Search to minimize the sum of both makespan and total weighted tardiness. Chen (2005), Wex et al. (2014), and Weng et al. (2001) developed problem-specific heuristics that minimize the total weighted completion time and the makespan.

For the parallel machine scheduling problem on unrelated machines with a total weighted completion time objective function, there are some approximation algorithms which are based on model relaxations and give a theoretical worst-case performance (see Li and Yang (2009) for an overview). However, these theoretical bounds are not promising for practical contexts as they are not very tight. Tabu search and genetic algorithms are used by Lin et al. (2011) and Vredeveld and Hurkens (2002) while Weng et al. (2001) and Wex et al. (2014) develop problem-specific heuristics. A good overview is further provided by Rodriguez et al. (2013).

## 2.4 High Performance Computing

HPC is used in many scientific disciplines, including biology, chemistry, physics, geology, weather forecasting, aerodynamic research, and computer science (Bell and Gray 2002; Vecchiola et al. 2009). But to the best of our knowledge, opportunities for modern parallel hardware architectures have been largely ignored in the scheduling literature.

Although HPC has become an integral part of several academic disciplines over the last decades, there is no commonly agreed definition of what HPC actually is. HPC implicitly refers to the use of parallel hardware architectures, which consist of a usually large set of interconnected multi-core processors (nodes). The particular relevance of parallel computing architectures lies in the limitation of speed improvement on a single core due to technological reasons (Hager and Wellein 2010).

At the extreme of HPC, supercomputers are used, which perform at or near the currently highest operational rate for computers. Today's 50 fastest supercomputers operate with more than 1 TFlop/s (TOP500 2014).

Parallel architectures provide a high potential to improve execution times of parallelizable algorithms. Dominating parallelization designs are OpenMP and MPI, with OpenMP allowing for intra-node shared-memory parallelization and MPI providing for inter-node distributed-memory network parallelization (MPI 2012; OpenMP 2013). OpenMP is used to execute a program using multiple threads (on multiple cores) and MPI is used to execute a program using multiple processes (on multiple processors). Both parallelization paradigms can be combined straightforwardly.

# 3 Scheduling Model and Parallel Depth-First Search Heuristic

In this section, we formulate the scheduling problem by means of a mathematical optimization model and we present a heuristic to obtain high-quality solutions. Finally, we present our parallelization concept for the proposed heuristic.

## 3.1 Problem Formulation

In this subsection, we provide a mathematical formulation of our scheduling problem $Rm/s_{ijk}, M_j/\sum w_j C_j$. It comprises jobs $j = 1, \ldots, n$ which have to be processed on exactly one machine $k = 1, \ldots, m$. Let $\Omega^k$ be the set of all feasible schedules on machine $k$, $a_{j\omega}$ be the number of times job $j$ is included in schedule $\omega$, and $c_\omega^k$ be the total weighted completion time of schedule $\omega$ on machine $k$. A decision variable $x_\omega^k$ is 1 if schedule $\omega$ is used on machine $k$ and 0 otherwise. The scheduling problem can then be formulated by the following binary linear optimization model.

$$\begin{aligned}
\min \quad & \sum_{k=1}^{m} \sum_{\omega \in \Omega^k} c_\omega^k \cdot x_\omega^k && \text{(BinLP)} \\
\text{s.t.} \quad & \sum_{k=1}^{m} \sum_{\omega \in \Omega^k} a_{j\omega} \cdot x_\omega^k = 1, \quad j = 1, \ldots, n \\
& \sum_{\omega \in \Omega^k} x_\omega^k \leq 1, \quad k = 1 \ldots, m \\
& x_\omega^k \in \{0,1\}, \quad k = 1, \ldots, m; \omega \in \Omega^k
\end{aligned}$$

Different formulations are proposed in the literature, see Li and Yang (2009) for an overview; the chosen one has proven to be particularly suitable for branch-and-price algorithms (Chen and Powell 1999; Chen and Powell 2003; Lopes and de Carvalho 2007; van den Akker et al. 1999).

## 3.2 Depth-First Search Heuristic

In this subsection, we describe a depth-first search (DFS) heuristic based on a branch-and-price (b&p) algorithm. A b&p algorithm is a branch-and-bound (b&b) algorithm with the special characteristic that the linear relaxation at each node of the b&b tree is solved using column generation, which was originally introduced to solve huge linear programs by Dantzig and Wolfe (1960). The original formulation of a b&p algorithm goes back to Barnhart et al. (1998).

B&p algorithms have been established to solve unrelated parallel machine scheduling problems exactly for rather small instances (Chen and Powell 1999; Lopes and de Carvalho 2007). Existing approaches use an implementation based on a node selection strategy where the active node with the best lower bound is selected to be explored next during the b&b algorithm. The selected node is then branched into two child nodes and nodes' relaxations are solved immediately afterwards. This strategy is called eager best-first strategy (Clausen and Perregaard 1999). However, these implementations fail at solving large instances to optimality due to time limitation.

| Line | Pseudo code |
|---|---|
| 1 | Solve root node's relaxation using column generation |
| 2 | **if** root node has integer optimal solution |
| 3 |     **return** the solution as optimal solution |
| 4 | **else** |
| 5 |     branch on root node by adding some restrictions |
| 6 |     initialize list of active nodes with both child nodes |
| 7 |     **while** no integer solution is found **do** |
| 8 |         select active node with highest depth to be explored next |
| 9 |         solve selected node's relaxation using column generation |
| 10 |         **if** selected node's relaxation is not infeasible |
| 11 |             branch on selected node by adding some restrictions |
| 12 |             add both child nodes to the list of active nodes |
| 13 | **return** integer solution as heuristic solution |

*Table 1. Pseudo code of the DFS heuristic*

We adapt these algorithms by implementing a different node selection strategy - the so called lazy DFS strategy - which selects the active node with the highest depth in the b&b tree to be processed next. After solving a node's relaxation, the node is branched into two child nodes which are added to the list of active problems. We terminate the procedure when an integer optimal solution for one of the subproblems is found. An integer optimal solution of an arbitrary subproblem is always a feasible (not necessarily optimal) solution of the original problem and therefore can be used as a heuristic solution value.

We select the lazy DFS strategy (i.e., branching on a node after solving its relaxation) based on two premises. First, the DFS strategy quickly finds a feasible solution for (BinLP) and second, lazy DFS strategies are more suitable for parallelization than eager DFS strategies (Clausen and Perregaard 1999).

Each node in the b&b tree is of a structure similar to (BinLP), only differing by some job ordering restrictions (depending on the branching decisions) which affect only the node-specific sets $\Omega^k$. The pseudo code of our DFS heuristic is presented in Table 1.

The code lines 2, 3, 4, 6, 7, 10, 12, and 13 are self-evident. We briefly explain the other code lines in the following.

The selection of an active node in line 8 is always possible since in the case of no active nodes, an optimal solution to the original problem would have been found, which means that the code would have already terminated the while loop.

The column (i.e., variable) generation procedure for solving the linear relaxation of a node (lines 1 and 9) is described in the following. As step 1 of the column generation procedure, a restricted form of the linear program is solved by considering only a (typically small) feasible subset of variables and setting the remaining variables to zero. An initial feasible subset of variables is either adopted from the solution of the parent node in the b&b tree or from a solution heuristic (for the root node). As step 2 of the column generation procedure, the algorithm determines whether there are any variables $x_\omega^k$ in the relaxed linear program that have a negative reduced cost $r_\omega^k$ which is given by

$$r_\omega^k = c_\omega^k - \sum_{j=1}^{n} a_{j\omega} \cdot \pi_j - \sigma_k$$

where $\pi_j, \sigma_k$ are the dual variables corresponding to the constraints in the relaxed linear program (Lopes and de Carvalho 2007). If there are variables with negative reduced cost, a fixed number of them are added to the restricted problem. Steps 1 and 2 of the column generation procedure are repeated until there are no variables left with negative reduced cost. Having reached this point, the optimal solution of the current restricted problem is also optimal for the node's linear relaxation, setting all remaining variables to 0 (Lopes and de Carvalho 2007). The problem of finding a variable with minimal reduced cost is called the pricing problem and will be discussed in the next subsection since it is highly suitable for parallelization.

The branching (lines 5 and 11) strategy is explained in the following. Let $\{x_\omega^k | k = 1, \ldots, m; \omega \in \Omega^k\}$ be the optimal solution of the linear relaxation of the root node (line 5) or the selected node (line 11) and let $\delta_{ij\omega}$ denote the number of times job i is processed immediately before job j in schedule $\omega$. Then we define the total flow of edge $(i, j, k)$ by

$$X_{ij}^k = \sum_{\omega \in \Omega^k} \delta_{ij\omega} \cdot x_\omega^k .$$

Branching on these variables has been proven to be very efficient in b&p algorithms (Chen and Powell 1999; Chen and Powell 2003; Lopes and de Carvalho 2007). We branch on the flow variable $X_{i^*j^*}^{k^*}$ with the largest integer infeasibility, i.e., the one closest to 0.5. The branching is conducted by modifying the sets of feasible schedules $\Omega^k$ in a way that, in one child node, $i^*$ is forced to be processed by $k^*$ immediately before $j^*$, and in the other child node, $i^*$ is forbidden to be processed by $k^*$ immediately before $j^*$ (Lopes and de Carvalho 2007).

### 3.3　Parallelization of the DFS Heuristic

There are two possible ways to utilize parallel computing for accelerating the execution speed of the DFS heuristic. First, the major part of the solution of each single b&b node can be parallelized on single multicore processors via shared-memory programming using OpenMP. Second, concurrently active nodes in the b&b tree can be processed independently using different multicore processors.

We describe the solution of the pricing problems in the column generation procedure (lines 1 and 9 in Table 1) and its parallelization using OpenMP in the following. Let T be an upper bound on the makespan, i.e., the time until all jobs have been processed, of an optimal solution for (BinLP). This value is unknown a priori but can be estimated efficiently (see Lopes and de Carvalho (2007) for details). For performance reasons, we allow schedules in $\Omega^k$ to be cyclic, i.e., each job is allowed to be processed more than once. Note that this does not affect the DFS heuristic solution because a cyclic schedule can never be used in an integer solution of a subproblem, since one of the coefficients $a_{j\omega}$ would be larger than 1.

Arbitrary jobs i, j on a machine k, require a time $p_j^k$ for processing j on k, a setup time $s_{ij}^k$ for processing j directly after i on k, and have a weight $w_j$. All of these parameters are assumed to be integers. We have a set $P_j^k$ of possible predecessors of job j on a machine k and a fictitious job 0 for modeling purposes. Let $f^k(j,t)$ denote the minimum reduced cost of all schedules $\omega \in \Omega^k$ that process j last and finish processing exactly at time t. We initialize $f^k(j,t) = +\infty$ for all $t \leq 0$, $f^k(0,0) := -\sigma_k$, and $f^k(0,t) = +\infty$ for all $0 \neq t \leq T$ (Lopes and de Carvalho 2007). The parallel pseudo code for the pricing problem is presented in Table 2. The omp parallel for pragmas divide the k-loop among all available threads using OpenMP.

| Line | Pseudo code |
|---|---|
| 1 | Initialize $f^k(j,t)$ as described above |
| 2 | **#pragma omp parallel for** |
| 3 |   **for** $k = 1, \ldots, m$ |
| 4 |     **for** $t = 1, \ldots, T$ |
| 5 |       **for** $j = 1, \ldots, n$ |
| 6 |         set $f^k(j,t) = \min_{i \in P_j^k} f^k(i, t - s_{ij}^k - p_j^k) + w_j t - \pi_j$ |
| 7 | **return** minimum reduced cost $\min_{k,j,t} f^k(j,t)$. |

*Table 2. Parallelized pricing problem*

The parallel pseudo code for the branching decision (code lines 5 and 11 in Table 1) is presented in Table 3.

Both code fragments in Table 2 and Table 3 are highly suitable for parallelization, since the work in each iteration of the k-loops can be done independently on different threads. Our pre-tests showed that the presented ordering of the for-loops and the selected parallelization strategy perform most efficient in terms of execution time since caching of data, i.e., making use of spatial and temporal data locality, is most efficient in this setting.

| Line | Pseudo code |
|---|---|
| 1 | **#pragma omp parallel for** |
| 2 |   **for** $k = 1, \ldots, m$ |
| 3 |     **for** $j = 1, \ldots, n$ |
| 4 |       **for** $i = 0, \ldots, n$ |
| 5 |         calculate $X_{ij}^k$ |
| 6 | **return** branching edge $(k^*, i^*, j^*) = \mathrm{argmin}_{k,i,j} |X_{ij}^k - 0.5|$ |

*Table 3. Parallelized branching decision*

In the following, we describe our strategy to parallelize the independent processing of concurrently active nodes of the b&b tree using the distributed-memory paradigm MPI. The MPI parallelizable part of the algorithm is represented by the while loop in code lines 7 to 12 in Table 1. We use a centralized master/worker setting for our parallelization as described in Clausen and Perregaard (1999) for instance. The pseudo code for the master process is presented in Table 4.

| Line | Pseudo code |
|---|---|
| 1 | **while** no integer solution is found **do** |
| 2 |   communicate with an idle worker process by |
| 3 |     receiving a solved node from the worker (if available) |
| 4 |   **if** received node's relaxation is not infeasible |
| 5 |     branching on selected node by adding some restrictions |
| 6 |     adding both child nodes to the list of active nodes |
| 7 |   selecting active node with highest depth to be explored next |
| 8 |   sending the selected node to the worker |

*Table 4. MPI master's pseudo code of while loop*

The pseudo code for the worker processes is illustrated in Table 5.

| Line | Pseudo code |
|---|---|
| 1 | **while** no integer solution is found **do** |
| 2 |   communicate with the master process by |
| 3 |     sending a solved node to the master (if available) |
| 4 |     receiving a new node from the master |
| 5 |   solve the received node's relaxation using column generation |

*Table 5. MPI workers' pseudo code of while loop*

The master process is responsible for the tree management while the worker processes perform the solutions of the nodes' relaxations - except the root node's relaxation, which is solved by the master process.

## 4 Computational Experiments

Our experiments were conducted on a Linux-based computing cluster consisting of multiple network-connected computing nodes. Each node is represented by a two-socket Intel Westmere X5675 shared-memory system with 6 cores per socket and a clock speed of 3.07 GHz. This is a standard architecture in modern high performance systems (Hager and Wellein 2010). The algorithm was coded in C++ and the restricted linear programs during column generation were solved via the GUROBI 6.0 C++ API.

We randomly generated ten instances for each of the instance sizes 300/300, 300/150, 300/100, 300/75, 300/60, 300/45, and 300/30, with n/m representing a setting with n jobs and m machines, in order to gain insights into how the DFS heuristic performs and how the parallelization works for different ratios of n to m.

The processing times in our experiments are uniformly distributed over $\{10, \ldots, 100\}$, the setup times (which are typically lower than the processing times) are uniformly distributed over $\{0, \ldots, 10\}$, and the weights are uniformly distributed over $\{1, \ldots, 10\}$. We generated 20 columns in each iteration of the column generation procedure and set the probability that a machine is eligible for processing a job to 20%. Similar settings have been used in the literature (Chen and Powell 1999; Chen and Powell 2003; Lopes and de Carvalho 2007; van den Akker et al. 1999; Weng et al. 2001).

## 5   Results and Discussion

In this section, we present and evaluate our results. We discuss the effectiveness of the DFS heuristic and its efficiency. Finally, we present our findings about the parallel implementation.

### 5.1   Effectiveness of the DFS Heuristic

For each instance, we calculated the solution of the DFS heuristic and of the best performing heuristic presented by Weng et al. (2001) and Wex et al. (2014) – we refer to this heuristic as SCHED – and documented execution times as well as solution quality (cf. subsection 2.3). The pseudo code of the SCHED heuristic is presented in Table 6. Let Cap denote the set of all pairs $(j,k)$ where machine $k$ is eligible of processing job $j$.

| Line | Pseudo code |
| --- | --- |
| 1 | Initialize the current completion time $c_k \coloneqq 0$, the current job $j_k \coloneqq 0$, and the current schedule $\sigma_k \coloneqq \emptyset$ for every machine $k$ |
| 2 | Initialize the set of remaining jobs $J \coloneqq \{1, \ldots, n\}$ |
| 3 | **while** $J \neq \emptyset$ **do** |
| 4 | set $(j^*, k^*) \coloneqq \mathrm{argmin}_{(j,k) \in Cap} \frac{c_k + s_{j_k j}^k + p_j^k}{w_j}$ |
| 5 | update $c_{k^*} \coloneqq c_{k^*} + s_{j_{k^*} j^*}^{k^*} + p_{j^*}^{k^*}$, $j_{k^*} \coloneqq j^*$, $\sigma_{k^*} \coloneqq \sigma_{k^*} \cup \{j^*\}$ |
| 6 | update $J \coloneqq J \setminus \{j^*\}$ |
| 7 | **return** $(\sigma_1, \ldots, \sigma_m)$ as the list of schedules |

*Table 6. Pseudo code of the SCHED heuristic*

The average gaps between the DFS heuristic solutions and (1) the best lower bounds (at time of algorithm termination) and (2) the solutions of the SCHED heuristic are presented in Table 7. These measures have been used in the literature, e.g., by Wex et al. (2014) and Schryen et al. (2015), and allow for quantifying the solution quality of DFS compared to (1) optimal solutions and (2) the SCHED heuristic.

| Instances | 300/300 | 300/150 | 300/100 | 300/75 | 300/60 | 300/45 | 300/30 |
| --- | --- | --- | --- | --- | --- | --- | --- |
| $Gap_{LB}$ | 0.00% | 0.03% | 0.11% | 0.10% | 0.18% | 0.36% | 0.35% |
| $Gap_{SCHED}$ | 3.26% | 4.99% | 5.27% | 6.18% | 7.68% | 8.83% | 11.35% |

*Table 7. Effectiveness of the DFS heuristic*

The gaps $Gap_{LB}$ are in fact upper bounds on the gap between the DFS heuristic solutions and the optimal solutions. A graphic illustration is provided in Figure 1.

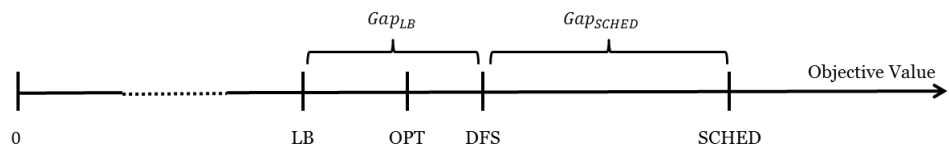

*Figure 1: Exemplary illustration of both gaps*

We conclude that the DFS heuristic solution values are either optimal or extremely close to the optimum as the average upper bound of the gap is not higher than 0.36% (scenario (300/45)). Even the maximum gap among all 70 tested instances was as low as 0.71%. Furthermore, the SCHED heuristic can be improved considerably by the DFS heuristic in terms of effectiveness.

## 5.2 Efficiency of the DFS Heuristic

The DFS heuristic calculated the solutions in less than 5 minutes even in a purely serial mode of execution. The average runtimes for each ratio n/m are presented in Figure 2.

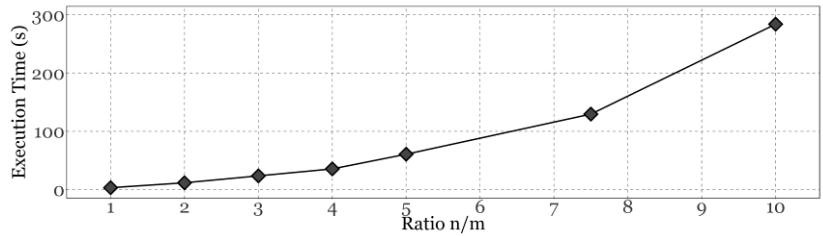

*Figure 2: Average runtimes of the DFS heuristic in serial execution mode*

While the runtimes of the DFS heuristic seem to increase super-linear, the runtimes of the SCHED heuristic were below one second in all tested instances. The super-linear increase of execution time of DFS is a disadvantage compared to SCHED. However, execution times are at an acceptable level for our tested instance sizes as the algorithm terminates within a few minutes.

## 5.3 Impact of HPC

The parallel implementation of our algorithm was tested on different configurations – 1, 6, and 12 threads on 1, 3, and 5 processes. This implementation is capable of substantially reducing runtimes for each of the instance sizes. Figure 3 depicts the decrease in execution time with an increasing number of threads on one process.

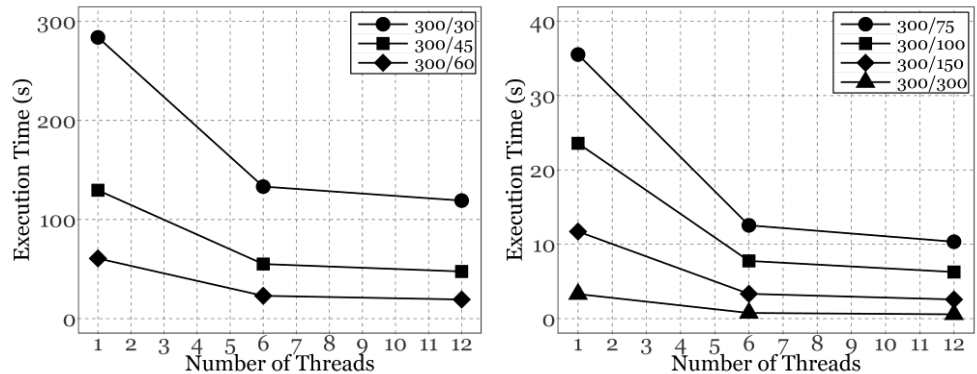

*Figure 3: Average runtimes of the DFS heuristic using multiple threads and one process*

The effect of HPC was evaluated using an established scalability metric – the comparison of theoretical and observed parallel speedup on one process and multiple threads (Hager and Wellein 2010). Let s denote the purely serial part of a program. The theoretical speedup using k threads can then be calculated as

$$S_k = \frac{1}{s + \frac{1-s}{k}}.$$

Note that $S_\infty = \lim_{k \to \infty} S_k = \frac{1}{s}$, which is known as Amdahl's law (Amdahl 1967). Table 8 lists the theoretical and observed speedups for all instance sizes and shows that our parallelization achieves runtimes which are just slightly below the theoretical boundaries.

The theoretical speedup decreases substantially with an increasing ratio of jobs to machines. The reason for this phenomenon is the increasing part of GUROBI's execution time compared to the overall execution time. Although GUROBI is claimed to use internal parallelization, we did not observe any parallelization effects in our context. Therefore, we pinned GUROBI to a single thread and consequently added GUROBI's execution time to the serial part of the algorithm.

| Type of SU # Threads | Theoretical 6 | Observed 6 | Theoretical 12 | Observed 12 | Theoretical ∞ |
|---|---|---|---|---|---|
| 300/300 | 4.51 | 4.36 | 6.33 | 5.65 | 12.28 |
| 300/150 | 3.55 | 3.54 | 4.76 | 4.62 | 7.27 |
| 300/100 | 3.05 | 3.04 | 3.84 | 3.77 | 5.19 |
| 300/75 | 2.87 | 2.84 | 3.53 | 3.44 | 4.59 |
| 300/60 | 2.70 | 2.65 | 3.25 | 3.17 | 4.09 |
| 300/45 | 2.42 | 2.36 | 2.83 | 2.74 | 3.39 |
| 300/30 | 2.19 | 2.14 | 2.49 | 2.40 | 2.88 |

*Table 8. Parallel OpenMP speedup (SU) of the DFS heuristic on one MPI process*

Our experiments further show that average execution times decrease by up to 2% in the best case but increase by up to 106% in the worst case using three or five processes. Using multiple processes therefore does not lead to an improvement in efficiency.

The reason for this phenomenon is the increasing number of explored nodes using multiple processes. The MPI approach enables the parallel processing of concurrently active nodes. This effects the exploration of unattractive nodes during the b&b algorithm and thus the DFS strategy becomes less efficient in our problem setting.

# 6   Conclusion

Many academic fields face scheduling problems as important decision making tasks. In this paper, we formulated a new heuristic for the NP-hard scheduling problem $Rm/s_{ijk}, M_j/\sum w_j C_j$. The heuristic is based on a branch-and-price algorithm and uses a lazy depth-first search strategy.

We showed that the proposed heuristic can solve large problem instances with 300 jobs in less than five minutes, even in serial execution of the algorithm. In addition, our heuristic returns solutions that differ only marginally from the optimal solution – 0.71% in the worst of all 70 tested scenarios.

The solutions were compared with an established solution heuristic for $Rm/s_{ijk}, M_j/\sum w_j C_j$ and the results show that this heuristic can be improved by an average 11.35% in the most difficult instance size using our approach. This leads to a high potential to save cost and time in practical applications.

Furthermore, we implemented different parallelization strategies for our heuristic and analysed their performance. We used shared-memory programming with OpenMP for the parallelization of the pricing problems and the branching decisions. We found that this approach substantially reduces runtimes of the heuristic when using multiple threads. In addition, we tested a distributed-memory MPI approach to process concurrently active nodes in the b&b tree independently on multiple processes. However, this approach tends to increase execution times because of an increasing number of explored b&b nodes. The best hardware setup for our algorithm would therefore be a single multicore processor with a large number of cores.

These findings provide substantial value for both the research community and practitioners. The research community can profit from our approach since we have bridged the largely unexplored gap between using HPC and solving scheduling problems. Practitioners can also benefit from our findings as we provide an easily accessible way to determine solutions of large-sized instances in reasonable time that are substantially better – in terms of the gap to the optimal solution - than solutions generated from an established heuristic proposed in the literature.

A limitation of our approach is the absence of real problem data. We coped with this problem by generating data sets randomly, which is a widely-used approach in the scheduling literature. However, we are in contact with emergency response organizations in order to evaluate our heuristic in real-world disaster response situations where a large number of incidents needs to be processed by rescue units. This is part of our ongoing work.

# 7　References